\documentstyle[12pt]{article}

\advance\textheight by \topskip
\newcommand{\Dir}{\kern -6.4pt\Big{/}}
\newcommand{\Dirin}{\kern -10.4pt\Big{/}\kern 4.4pt}
\newcommand{\DDir}{\kern -7.6pt\Big{/}}
\newcommand{\DGir}{\kern -6.0pt\Big{/}}
\newcommand{\ar}{\rightarrow}
\newcommand{\be}{\begin{equation}}
\newcommand{\ee}{\end{equation}}

\begin{document}

\thispagestyle{empty}
\setcounter{page}{0}

\begin{flushright}
{\large DFTT 37/95}\\
{\rm June 1995\hspace*{.5 truecm}}\\
\end{flushright}

\vspace*{\fill}

\begin{center}
{\Large \bf Six--jet decay of off--shell WW pairs at
$e^+ e^-$ colliders \footnote{ Work supported in part by Ministero
dell' Universit\`a e della Ricerca Scientifica. \hfill\break\indent
\ e--mail: accomando,ballestrero,maina@to.infn.it}}\\[2cm]
{\large Elena Accomando, Alessandro Ballestrero and Ezio Maina}\\[.3 cm]
{\it Dipartimento di Fisica Teorica, Universit\`a di Torino}\\
{\it and INFN, Sezione di Torino}\\
{\it v. Giuria 1, 10125 Torino, Italy.}\\
\end{center}



\vspace*{\fill}

\begin{abstract}
{\normalsize
Six--jet events via WW pairs, $e^+e^- \ar W^+W^- \ar q_1
\overline{q_1} q_2 \overline{q_2} g g$
are studied at tree level using helicity amplitudes.
This is the dominant production mechanism for six--jet final states
at Lep II energy. ISR effects are taken into account.
Total production rates
as a function of $y_{cut}$ are given. The relevance of these processes
for the issue of colour reconnection is discussed. The cross section
for five--jet production via WW pairs at Lep II is also presented.
}
\end{abstract}

\vspace*{\fill}

\newpage
\subsection*{Introduction}
Several different mechanisms contribute to six--jet production. At
Lep II energies they fall into two broad classes:
the `point--like' annihilation of $e^+e^-$ to $Z^0,\gamma$,
which dominates at LEP I energy, and vector boson pair production
whose importance grows with increasing energy and
dominates above the $WW$ treshold. At still larger energy one should
also consider the contribution of three vector bosons.
Interference effects
between the two set of diagrams are expected to be small
since the point--like contribution is dominated by the radiation of
four gluons from the two primary quarks with only a small fraction
of events with four and possibly six quarks. On the contrary
in WW decays there are
at least four quarks. Moreover only half of the WW events
are flavour neutral and can therefore interfere with the point--like
contribution.
\par
In this paper we study the production of off--shell WW pairs and their
subsequent decay to five and six jets at $e^+e^-$ colliders.
Since $\alpha_s (M_Z) \approx .12$, gluons are radiated with high probability
and five-- and six--jet fractions are large also
at experimentally relevant values of $y_{cut}$. This mechanism is the main
source of five-- and six--jet events with total hadronic mass much greater
than the mass of the $Z$. A large number of events with total hadronic mass
close to $M_Z$ is produced by radiative return to the $Z$ peak via
initial state radiation.
\par
One of the main goals of Lep II is the measurement of the W mass
with high accuracy. Fully hadronic decays of the $WW$ pairs represent
4/9 of the cross section.
Apart from all the usual uncertainties
related to measuring jet energies and directions, which can be
partially eliminated by using energy--momentum conservation and possibly
the approximate equality of the $W^+$ and $W^-$ masses,
there are uncertainties which
stem from the fact that two decays occur in the same event.
A jet from, say , the decay of the $W^+$ can be closer, in a given
reconstruction scheme, to a jet from the decay of the $W^-$ than to
any other jet from the positively charged $W$. If the stray jet carries
large energy, then
the reconstructed masses will be quite far from the $W$ mass which is
already at present known
with an error of about 200 MeV.
One could imagine that such events could be discarded.
However, due to the intrinsic width of the $W$ and to all
experimental uncertainties, it is impossible to impose very stringent
cuts on the difference between the measured mass and the true mass.
It is therefore important to estimate the effect of misassigned jets,
including the large angle emission tails. For this purpose
it is well known that matrix element calculations describe the full
angular distribution of jets better than parton shower models while
the latter are superior for small angle radiation.
It should be mentioned that, due to
the number and complexity of the diagrams involved, it is unlikely that
a full ${\cal O}(\alpha_s^2)$ calculation will be produced in the
foreseeable future, and only tree--level matrix elements will be
available.
\par
Recently it has been pointed out \cite{gpz, sjo_kho, gu_ha}
that cross--talk between the decay of the two $W$'s
can take place also at energy scales much smaller than those typical
of jets. This destroys the notion of
two separate decays and cast doubts on our ability to reconstruct
the masses of the original sources from the decay products,
producing potentially large uncertainties.
The perturbative contribution to the phenomenon, is related,
in the Feynman diagram language, to the presence in the expression
for the cross section of interference terms, in which real or virtual
partons emitted in the decay of one $W$
are reabsorbed by the decay products of the oppositely charged $W$.
One of the simplest example is given by the interference between
diagrams of type $A_1$ and diagrams of type $A_2$ or between
diagrams of type $A_3$ and diagrams of type $A_4$ in fig. 1.\par
The contribution of hard gluons, which can be treated
perturbatively, to colour rearrangement effects is expected to be
small.
The argument can be summarized as follows. Only interference
terms between different diagrams can produce colour rearrangement;
terms in which the gluons emitted by one line are reabsorbed by the
same line do not mix the original decays.
It is easily realized
that all diagrams with a single gluon exchange between the two fermion lines
originating from different $W$'s vanish when the sum over colours is performed.
When two, real or virtual, gluons are exchanged, on the contrary non--zero
contributions are possible. As a consequence colour rearranged final states are
suppressed by $\alpha_s^2$ compared with normal ones. Moreover, it can
be shown that the color factor for the contributions where color rearrangement
takes place is smaller by a factor $(N_c^2 -1)^{-1}$ with respect to the color
factor of the leading terms in $1/N_c$ at the same order in $\alpha_s$.
If one compares these terms with the lowest order cross section the ratio
of colour factors is $\frac{1}{4}\frac{(N_c^2 -1)}{N_c^2} = \frac{2}{9}$.
A further suppression of the hard gluon contribution
to colour reconnection derives from
the finite width of the $W$'s. As an example consider the interference between
the diagrams of set $A_1$ and the diagrams of set $A_2$ in fig. 1.
If the gluons carry
a substantial amount of energy at least two of the $W$ propagators must be
far off mass shell and therefore the contribution of this term is negligible.
The authors of ref. \cite{sjo_kho} give the following estimate for the ratio
between the contribution of colour rearranged diagrams and the total cross
section:
\begin{equation}\label{deltasigma}
\frac{\Delta \sigma}{\sigma} \le \left(
       \frac{C_F \alpha_s}{N_c} \right)^2
             \frac{\Gamma_W}{M_W}
\end{equation}
In this expression
$\alpha_s$ is to be evaluated at $Q^2 = \Gamma_W^2$ where
$\alpha_s \approx .25$ so that $\Delta \sigma / \sigma \le 10^{-3}$.
\par
The non--perturbative contribution to colour reconnection
can be described in very simple terms in the string model, arguably the
best tool for understanding the fragmentation and hadronization phase
of jet development, which links the perturbative gluon cascade to
the observed hadrons. In this model, particle production results
from the iterative splitting of color strings or antennae which are
color singlets and evolve independently of each other.
When two color singlet sources, like the two $W$'s  at Lep II, decay close
to each other in space--time there two are different ways of forming colour
singlets out of the two  $q \bar q$ pairs. If the primary decays are
$W^+ \ar q_1 \overline{q_2}$ and $W^- \ar q_3 \overline{q_4}$ \ the
$q_1 \overline{q_2}$ and $q_3 \overline{q_4}$ pairs are always in a relative
singlet state. However, with probability 1/9, the
$q_1 \overline{q_4}$ and $q_3 \overline{q_3}$ pairs can also form
colour singlet states.
Since the evolution of the antennae is determined by their
invariant mass the multiplicity and distribution of detected particles
depend strongly on the pairing pattern which is selected.\par
A more quantitative estimate of color reconnection effects can be
obtained using JETSET \cite{jetset}.
In JETSET the primary decay is followed by a perturbative
shower evolution, in which successive emissions are strongly
ordered in angle as dictated by colour coherence. The showers
develop as a Markov--chain process and no colour reconnection takes
place in this phase. The two showers expand and when the
hadronization phase begins they can overlap. During fragmentation
each gluon acts as a $q \bar q$ pair and real hadrons are produced
by the recursive splitting of colour strings extending between $q \bar q$
pairs. It is at this non--perturbative
stage that the authors of ref. \cite{sjo_kho} allow
colour strings to interact,
to break and form new strings which may join quark originating from
different showers, producing particles which cannot be assigned unambigously
to any of the original $W$'s. The details
of the reconnection mechanism are unknown and require
some additional modeling with respect to
the well tested basic rules used in JETSET.
It is this lack of information in the choice of the reconnection mechanism,
that produce the spread in magnitude and even in sign of
the predictions for the effects of colour reconnection in the
measurement of $M_W$.

\subsection*{Results}
We have studied the reaction $e^+e^- \ar W^+W^- \ar q_1
\overline{q_1} q_2 \overline{q_2} g g$ at tree level.
Six quark final states have not been included since, at $e^+e^-$
colliders, when
a pair of gluons is replaced by a quark--antiquark pair, summed over all
allowed flavour combinations, the cross section typically decreases
by an order of magnitude.
Quark masses have been neglected since the contribution of $b$--quarks
is severely suppressed by the smallness of the $V_{bc}$ element of
the CKM matrix. The relevant diagrams are shown schematically in fig. 1.
For simplicity the lepton part of the diagrams is not drawn
and a sum over the three possible structures which describe
$e^+e^-\rightarrow WW$ is understood.
The set labelled A1 (A2) includes the twenty--four
diagrams in which both gluons
are emitted in the decay of the $W^+$ ($W^-$). Analogously, in A3
we include the twelve diagrams in which one gluon, let us call it $g_5$
is emitted in the decay of the $W^+$ while the second, $g_6$, is emitted
in the decay of the $W^-$. In A4 the positions of the two gluons are
interchanged.
\par
All matrix elements have been computed using the formalism
presented in ref. \cite{method}
with the help of a set of routines, called  PHACT \cite{phact}, which
generate the building blocks of the helicity amplitudes
semi--automatically.
In our experience this method is faster than
others which are commonly used \cite{ks, mana, hz}.
Further gains in speed can be obtained
avoiding subroutine and function calls. In order to achieve
this goal the routines in PHACT instead of computing the numerical values
of the different terms, write the corresponding FORTRAN code, which can then
be compiled and run. In the formalism of
\cite{method} it is easy to save
every sub--diagram and then to reuse it several times.
With this procedure we have generated a rather large piece of code,
which however runs quite fast, and therefeore
can be used in high statistics Montecarlo runs.
\par
The amplitudes have been checked for QCD gauge invariance.
We have used  $M_Z=91.1$ GeV, $\Gamma_Z=2.5$ GeV, $M_W=80.6$ GeV,
$\Gamma_W=2.06$ GeV,
$\sin^2 (\theta_W)=.23$, $m_b=5.$ GeV, $\alpha_{em}= 1/128$ and
$\alpha_{s}= .115$ in the numerical part of our work.
\par
Our main results are presented in fig. 2 through 4. Fig. 2a shows
the total cross section as a function of $y_{cut}$ in the JADE
\cite{jade} scheme
for the three energies which have been agreed on or are under consideration
for Lep II, that is $\sqrt{s} = 175,\  190$ and $205$ GeV.
At such energies the typical value of $y_{cut}$ we have studied, namely
$y_{cut} = 5\times 10^{-3}$ corresponds to a jet--jet invariant mass
above 12 GeV. In Fig. 2a initial state radiation (ISR) has been neglected.
We see that for
a fixed value of $y_{cut}$  the cross section decreases with increasing
energy, while, in this range, the $WW$ cross section is increasing.
This is due to the definition of $y_{cut}$:
\be
y_{cut} = \frac{2 E_i E_j}{s} (1 - \cos \theta_{ij} )
\ee
which means that at larger center of mass energy larger jet--jet invariant
mass are required for an event to pass the cut. On the other hand the mass
scale of the decays is obviously given by the $W$ mass and does
not change with $s$. Since small invariant masses are more likely for two jets
from the same $W$ that for jets from different $W$'s one indeed expects
the fixed--$y_{cut}$ cross section not to grow as quickly as the
cross section for $e^+e^- \ar WW$.\par
In Fig. 2b we study ISR effects for six--jet production
at $\sqrt{s} = 175$ and $\sqrt{s} = 205$. The results at $\sqrt{s} = 190$
have been computed but are not shown for the sake of clarity.
We have used the
structure function approach at leading--log as in ref. \cite{pass_pavia}.
The cross section is decreased by about 20\%
at $\sqrt{s} = 175$ and by about 10\% at $\sqrt{s} = 205$. The effect at
$\sqrt{s} = 190$ nicely interpolates those at the two energies which are shown.
\par
For comparison in Fig. 3 we present the cross sections
for five jet production at the same energies used for six jets.
We have repeated the calculation of ref. \cite{brown} and found complete
agreement.
The results without ISR are given in fig. 3a and those which compare
results with and without ISR effects in fig. 2b.
The dot--dashed line in fig. 3a shows the five jet background
cross section \cite{fgk, our5} from $e^+e^- \ar  q \bar q ggg$ and
$e^+e^-  \ar q_1 \overline{q_1} q_2 \overline{q_2} g$.
\par
There is an aspect which deserves some comment.
When our results for five--jet production \cite{our5}
at Lep I are compared with the data
presented by ALEPH \cite{expMC} and OPAL \cite{alphas}
it is clear that the absolute normalization is about a factor of five
too small. The simplest explanation for this discrepancy is
our choice for $\alpha_s$.
In fact we have used $\alpha_s =.115$ which corresponds to
$Q^2= M^2_{z^0}$ with $\Lambda_{\overline{MS}}=200$ MeV with five
active flavours in the standard formula:
\begin{equation}
\alpha_s(Q) = {{1}\over{b_0 \log (Q^2/\Lambda^2)}} \left[
1- {{b_1 \log (\log (Q^2/\Lambda^2))}\over{b_0^2 \log (Q^2/\Lambda^2) }}
\right]
\end{equation}
The analysis of shape variables and jet rates to ${\cal O}(\alpha_s^2)$
has shown that, in order to get agreement between the data and the
theoretical predictions, the scale of the strong coupling constant
has to be chosen to be $Q= x_\mu M_{Z^0}$, with $x_\mu \approx 0.1$
\cite{alphas}.
It has later been shown that when the relevant logarithms are properly
resummed \cite{catani} agreement is obtained for much larger
values of the scale, $x_\mu \approx 1.$ \cite{resummed}.
It is therefore not surprising that our tree level expressions require
a relatively small scale in order to describe the data.
As a consequence the cross sections presented in Fig. 2a,b are expected
to be somewhat underestimated. Since $\alpha_s$ is an
overall factor our results
can be easily modified if a different value for the strong coupling
constant is preferred.
\par
Leaving aside the issue of best choice of $\alpha_s$,
with an expected luminosity between 300 and 500 $pb^{-1}$ it is clear
that only for $y_{cut} \le 1.5 \times 10^{-2}$ six jet events can be observed.
The cross section grows very rapidly and for $y_{cut} = 5 \times 10^{-3}$
one expects ${\cal O} (100)$ decays of a $WW$ pair to six jets
per year and per experiment.
\par
In Fig. 4 we present the gluon energy spectrum. The continous line
gives the differential distribution for one fixed gluon, while the dashed line
and the short--dash--long--dash line give the energy spectrum for the
most and least energetic gluon respectively. These quantities are not directly
observable but since in most cases the softest jets are the gluon ones they
give an indication of what sort of low energy jets one might expect in $WW$
decays. The softest gluon energy distribution is rather narrow
and peaks at about 8 GeV. The most energetic gluon has a much broader
distribution with a maximum at about 20 GeV.
\par
In order to obtain a quantitative estimate of the colour reconnection
contribution to the cross section we have integrated over phase space
the interference between the diagrams in the set A1 and those in A2 and
the interference between the diagrams in the set A3 and those in A4.
The interference between (A1+A2) and (A3+A4) is zero.  In the following we
will call the result $\sigma_{int}$.
These integrals are
particularly challenging. The magnitude and phase of the integrand
change rapidly at all $W$ poles and since different sets of particle
reconstruct the $W$ mass in different diagrams it is
impossible to eliminate the strong peaking structure with the standard
change of variable $k_i^2- M_W^2 = M_W \Gamma_W \tan \theta_i$.
In addition
the requirement that all jet--jet invariant masses be larger than
$M_{min} = s y_{cut}$ introduces discontinuities
within the integration region. The accuracy of our results for the
interference terms is typically of the order of 1$\div$2\%, the least accurate
point being at $y_{cut} = 1 \times 10^{-3}$ and
$\Gamma_W = 2.06$ GeV where the error is about 6.6\%. The results presented in
fig. 2 through 4 have an accuracy well below 1\%.\par
In the narrow width limit the cross section scales as $(\Gamma_W)^{-2}$
and the bound (\ref{deltasigma}) suggests that the interference terms are
proportional to $(\Gamma_W)^{-1}$. Therefore we have investigated the
behaviour of $\sigma_{int}$ as a function of the $W$ width in the range
between 10 and 2 GeV. Numerical instabilities become less severe
if one artificially increases
the $W$--boson width and this allows us to keep the numerical
accuracy of the integration under control.
Our result are presented in Fig. 5 for $\sqrt{s} = 175 \ GeV$ and two
$y_{cut}$ values,
$y_{cut} = 6.5\times 10^{-3}$ and $y_{cut} = 1\times 10^{-3}$.
For comparison we also plot the curves of the form $y= a/\Gamma_W + b$
which interpolate our results at $\Gamma_W = 10$ GeV and $\Gamma_W = 7$ GeV.
In events which do contribute to $\sigma_{int}$ the gluon energies are
restricted from below by the cut on invariant masses and from above by
$W$--width effects as explained previously. The resulting band becomes narrower
for increasing $y_{cut}$ and for decreasing $\Gamma_W$.
This can be clearly seen in Fig. 5. For large values of the $W$--width the
results approximately behave as $a/\Gamma_W + b$ while for values of
$\Gamma_W$ approaching the physical value they gradually fall
below the reference curve. This is more evident for $y_{cut} = 6.5\times
10^{-3}$ where the lower limit of the allowed band is higher.
Taking the cross section for $e^+e^- \ar W^+W^-
\ar q_1 \overline{q_1} q_2 \overline{q_2}$ to be 8 $pb$ at
$\sqrt{s} = 175 \ GeV$, $\Delta \sigma /\sigma$ turns out to be
$1 \times 10^{-5}$ at $y_{cut} = 6.5 \times 10^{-3}$ and
$5 \times 10^{-4}$ at $y_{cut} = 1 \times 10^{-3}$, within the bound
of ref. \cite{sjo_kho}. The ratio of the interference terms with
the six--jet cross section has a milder dependence on $y_{cut}$,
being $3.3 \times 10^{-4}$ at $y_{cut} = 6.5 \times 10^{-3}$ and
$5.7 \times 10^{-4}$ at $y_{cut} = 1 \times 10^{-3}$.

\subsection*{Conclusions}
We have computed at tree level the cross section for the process
$e^+e^- \ar W^+W^- \ar q_1 \overline{q_1} q_2 \overline{q_2} g g$
which is the dominant contribution to six--jet production at Lep II energies.
With a luminosity of 500 $pb^{-1}$
one expects ${\cal O} (100)$ decays of a $WW$ pair to six jets
per year and per experiment for $y_{cut} = 5 \times 10^{-3}$.
We have studied the lowest order non--trivial perturbative contribution to
colour reconnection. In the range we have considered, namely
$y > 1 \times 10^{-3}$, it is small and within the bound of ref.
\cite{sjo_kho}.
We have presented the cross sections for the process
$e^+e^- \ar W^+W^- \ar q_1 \overline{q_1} q_2 \overline{q_2} g$ at Lep II.


\newpage

\subsection*{Figure Captions}

\begin{description}

\item[Fig. 1] Representative diagrams of the decay part
contributing to $e^+e^-\rightarrow
WW \rightarrow q_1\overline q_2 q_3\overline q_4 gg$.
For simplicity the lepton part of the diagrams is not drawn
and a sum over the three possible structures which describe
$e^+e^-\rightarrow WW$ is understood.
In set A1 (A2) we include the eight diagrams in which both gluons
are emitted in the decay of the $W^+$ ($W^-$).
Analogously in set A3 we include the four diagrams in which
gluon  $g_5$
is emitted in the decay of the $W^+$ while $g_6$ is emitted
in the decay of the $W^-$. In A4 the positions of the two gluons
are interchanged.

\item[Fig. 2] Total cross section  as a
function of $y_{cut}$ in the JADE scheme
for six--jet production via $WW$ at $\sqrt{s} = 175$ GeV (continuous line),
$\sqrt{s} = 190$ GeV (dashed line) and $\sqrt{s} = 205$ GeV (dotted line).
In Fig. 2a ISR effects are not included.
In Fig. 2b we compare results which include ISR effects with
those without ISR, as explained in the main text,
at $\sqrt{s} = 175$ GeV and $\sqrt{s} = 205$ GeV.

\item[Fig. 3] Total cross section  as a
function of $y_{cut}$ in the JADE scheme
for five--jet production via $WW$ at $\sqrt{s} = 175$ GeV (continuous line),
$\sqrt{s} = 190$ GeV (dashed line) and $\sqrt{s} = 205$ GeV (dotted line).
The dot-dashed line in Fig. 3a
gives the QCD contribution to five--jet production
through point--like annihilation at  $\sqrt{s} = 175$ GeV.
In Fig. 3a ISR effects are not included.
In Fig. 3b we we compare results which include ISR effects with
those without ISR, as explained in the main text,
at $\sqrt{s} = 175$ GeV and $\sqrt{s} = 205$ GeV for
the $WW$ signal only.

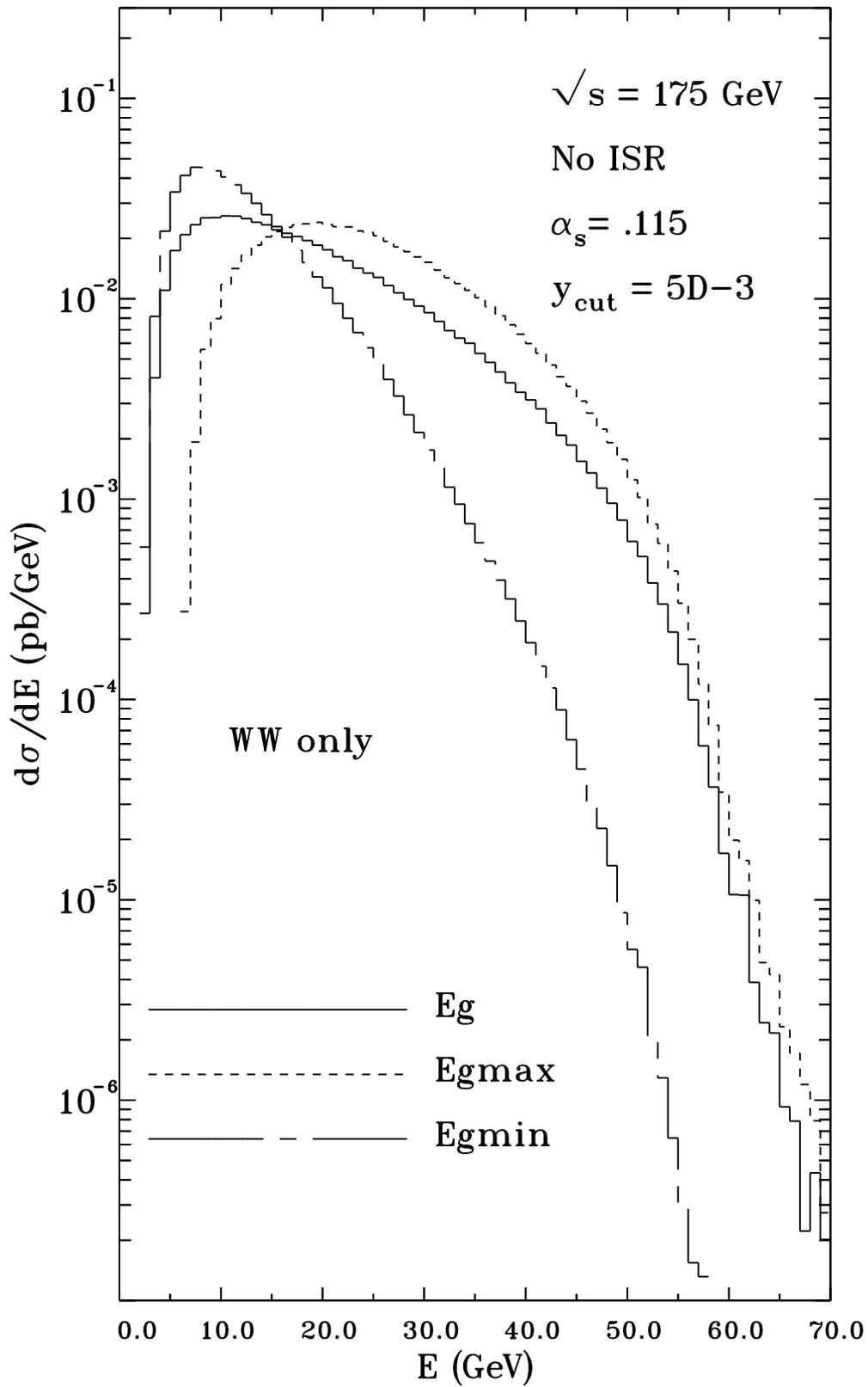
\item[Fig. 4] Gluon spectra in six--jet production
via $WW$ at $\sqrt{s} = 175$ GeV. The continous line gives
the energy distribution for a single gluon, while the short--dashed
line and the long--dash--short--dash line give the spectrun of the most
and least energetic of the two gluons, respectively.

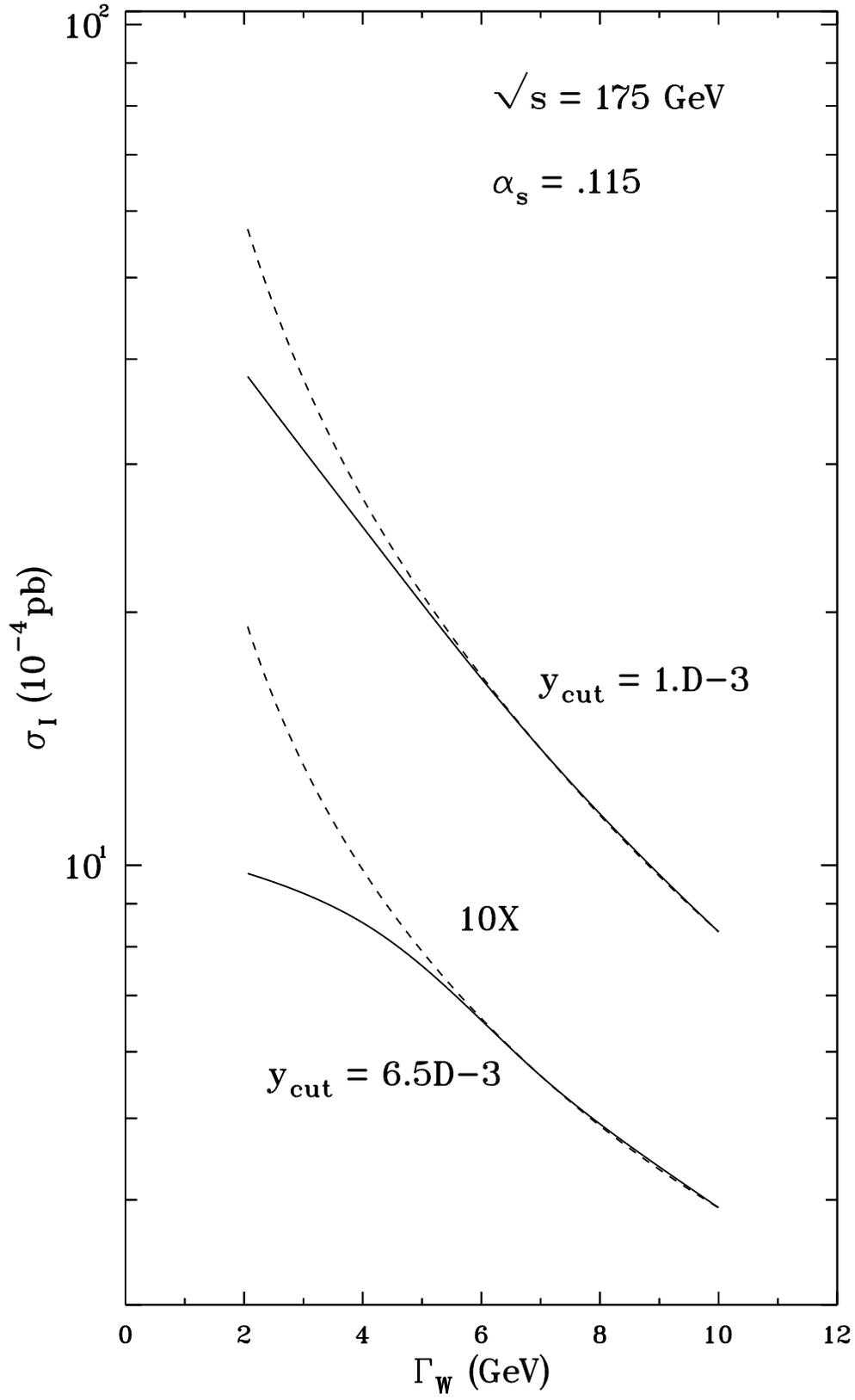
\item[Fig. 5] Integral over phase--space of the sum of the interference
between the diagrams in set A1 with those in set A2 and of the diagrams
in set A3 with those in set A4 at $\sqrt{s} = 175$ GeV.

\end{description}

\end{document}